\documentclass[11pt,fleqn,twoside]{article}
\usepackage{latexsym}
\makeatletter
\newcommand{\prava}{\footnotesize\it
\begin{flushright}
\begin{minipage}{6cm}
Copyright \copyright 1998 by D.L. Blackmore, Y.A. Prykarpatsky and
R.V. Samulyak
\end{minipage}
\end{flushright}}

\newcommand{\name}[1]{\begin{flushleft}
                       \LARGE \bf #1
                       \end{flushleft}\vspace{-3mm}}

\newcommand{\Author}[1]{\begin{flushleft}
                       \it #1 \end{flushleft}}

\newcommand{\Adress}[1]{\begin{flushleft}
                       \it #1 \end{flushleft}}

\newcommand{\Date}[1]{\begin{flushleft}
                      \small  \it #1 \end{flushleft}}

\newcommand{\ehkol}{Author \ name}
\newcommand{\ohkol}{Article \ name}
\renewcommand{\@evenhead}{
\hspace*{-3pt}\raisebox{-15pt}[\headheight][0pt]{\vbox{\hbox to \textwidth
{\thepage \hfil \ehkol}\vskip4pt \hrule}}}
\renewcommand{\@oddhead}{
\hspace*{-3pt}\raisebox{-15pt}[\headheight][0pt]{\vbox{\hbox to \textwidth
{\ohkol \hfil \thepage}\vskip4pt\hrule}}}
\renewcommand{\@evenfoot}{}
\renewcommand{\@oddfoot}{}

\newcommand{\be}{\begin{equation}}
\newcommand{\ee}{\end{equation}}
\newcommand{\ba}{\hspace*{-5pt}\begin{array}}
\newcommand{\ea}{\end{array}}

\newcommand{\ds}{\displaystyle}
\makeatother

\begin{document}
\setcounter{page}{54}

\thispagestyle{empty}

\renewcommand{\ehkol}{D.L. Blackmore, Y.A. Prykarpatsky
and R.V. Samulyak}
\renewcommand{\ohkol}{Lie-Invariant Geometric Objects}

\begin{flushleft}
\footnotesize \sf
Journal of Nonlinear Mathematical Physics \qquad 1998, V.5, N~1,\ 
\pageref{black-fp}--\pageref{black-lp}.\hfill {\sc Article}
\end{flushleft}

\vspace{-5mm}

{\renewcommand{\footnoterule}{}
{\renewcommand{\thefootnote}{}  \footnote{\prava}}

\name{The Integrability of Lie-invariant Geometric Objects Generated
by Ideals in the Grassmann Algebra}\label{black-fp}

\Author{D.L. BLACKMORE~$^\dag$, Y.A. PRYKARPATSKY~$^{\ddag}$ and
R.V. SAMULYAK~$^{\ddag}$}


\Adress{$^\dag$
\ Department of Mathematics and Center of Applied Mathematics and\\
$\phantom{^{\ddag }}$~Statistics at the New Jersey Institute of Technology, Newark NJ 07102
USA\\[1mm]
$^{\ddag}$
Department of Nonlinear Mathematical Analysis at the Institute of Applied
Problems \\
$\phantom{^{\ddag}}$~of Mechanics and Mathematics, NAS, L'viv 290601, Ukraina}

\Date{Recieved September 17, 1996; Accepted November 12, 1997}

\begin{abstract}
\noindent
We investigate closed ideals in the Grassmann algebra serving as bases
of Lie-invariant geometric objects studied before by E.Cartan.
Especially, the E.Cartan theory is enlarged for Lax integrable
nonlinear dynamical systems to be treated in the frame work of the
Wahlquist Estabrook prolongation structures on jet-manifolds and
Cartan-Ehresmann connection theory on f\/ibered spaces.
General structure of integrable one-forms augmenting the two-forms
associated with a closed ideal in the Grassmann algebra is studied
in great detail. An ef\/fective Maurer-Cartan one-forms construction
is suggested that is very useful for applications. As an example of
application the developed Lie-invariant geometric object theory for
the Burgers nonlinear dynamical system is considered having given
rise to f\/inding an explicit form of the associated
Lax type representation.
\end{abstract}

\section{General setting}

It is well known [1, 4] that motion planning, numerically controlled
machining and robotics are just a few of many areas of manufacturing
automation in which the analysis and representation of swept volumes plays
a crucial role. The swept volume modeling is also an important part of
task-oriented robot motion planning. A typical
motion planning problem consists in  a collection of objects moving
around obstacles from an initial to a f\/inal conf\/iguration. This may  include
in particular, solving the collision detecting problem.

When a solid object undergoes a rigid motion, the totality of points through
which it passed constitutes a region in space called the swept volume. To
describe the geometrical structure of the swept volume we pose this
problem as one of geometric study of some manifold swept by surface
points using powerful tools from both modern dif\/ferential geometry and
nonlinear dynamical systems theory [2-4, 7, 8] on manifolds. For some
special cases of the euclidean motion in the space ${\bf R}^3$ one can
construct a very rich hydrodynamic system [1] modelling a sweep
f\/low, which appears to be a completely integrable Hamiltonian system
having a special Lax type representation. To describe in
detail these and other properties of swept volume dynamical systems in this
article we develop dif\/ferential-geometric Cartan's theory of Lie-invariant
geometric objects generated by closed ideals in the Grassmann algebra as well
as investigate some special examples of euclidean motions in ${\bf R}^3$
leading to Lax type integrable dynamical systems on functional manifolds.

Let a Lie group $G$ act on an analytical manifold $Y$ in the transitive
way, that is the action $G\times Y\stackrel{\rho}{\to}Y$
generates some nonlinear exact
representation of the Lie group $G$ on the manifold $Y$. In the frame of the
Cartan's dif\/ferential geometric theory, the representation
$G\times Y\stackrel{\rho}{\to} Y$
can be described by means of a system of dif\/ferential 1-forms
\begin{equation}
\bar\beta ^j:=dy^j+\sum^r_{i=1}\xi^j_i\bar\omega^i(a;da)\in\Lambda^1(Y\times G)
\end{equation}
in the Grassmann algebra $\Lambda(Y\times G)$ on the product $Y\times G$,
where
$\bar\omega^i(a;da)\in T^*_a(G)$,
$i=\overline{1,r=\mbox{dim}\,G}$ is a basis of left
invariant Cartan's forms of the Lie group $G$ at a point $a\in G$,
$y:=\{y^j: \,j=\overline{1,n=\mbox{dim}\,Y}\}\in Y$
and $\xi^j_i:\,Y\times G\to {\bf R}$
are some smooth real valued functions. The following Cartan's theorem is basic
in describing  a geometric object invariant with respect to the mentioned above
group action $G\times Y\stackrel{\rho}{\to}Y$:

}

\medskip

\noindent
{\bf Theorem 1.}(E.Cartan). {\it The system of dif\/ferential forms (1) is  a
system of an invariant geometric
object if and only if the following conditions are fulf\/illed:

i) the coef\/f\/icients $\xi^j_i\in C^{\infty}(Y;{\bf R})$ for all $i=\overline{1,r},\,
j=\overline{1,n}$, are some analytical functions on $Y$;

ii) the dif\/ferential sysetm (1) is completely integrable within the
Frobenius-Cartan criterium.}

\medskip

The Theorem 1 says that the dif\/ferential system (1) can be written down as
\begin{equation}
\bar\beta ^j:=dy^j+\sum^r_{i=1}\xi^j_i(y)\bar\omega^i(a;da),
\end{equation}
where one-forms  $\{\bar\omega^i(a;da): \,i=\overline{1,r}\}$ satisfy the
standard Maurer-Cartan equations
\begin{equation}
\bar\Omega^j:=d\bar\omega^j+\frac 12\sum^r_{i,k=1}c^j_{ik}\bar\omega^i\land
\bar\omega^k:=0
\end{equation}
for all $j=\overline{1,r}$ on $G$, coef\/f\/icients $c^j_{ik}\in{\bf R}, \,
i,j,k=\overline{1,r}$, being the corresponding structure constants of the
Lie algebra $\cal G$ of the Lie group $G$.

Let us consider here a case when the set of canonical Maurer-Cartan one-forms
\linebreak $\{\bar\omega^i(a;da)\in T^*_a(G):\,i=\overline{1,r}\}$ is def\/ined via the
scheme:
\begin{equation}
\begin{array}{c@{}c@{}c@{}c@{}c}
T^*(M\times Y) & \stackrel{s^*}{\longrightarrow} & T^*(\bar M) &
\stackrel{\mu^*}{\longleftarrow} & T^*(G\times Y) \\
\downarrow & \  & \downarrow & \  & \downarrow \\
M\times Y & \stackrel{s}{\longleftarrow} & \bar M &
\stackrel{\mu}{\longrightarrow} &
G\times Y
\end{array}
\end{equation}
where $M$ is a given smooth f\/inite-dimensional manifold with some submanifold
$\bar M\subset M$ imbedded into it as $s:\bar M\to M\times Y$, and $\mu:\,
\bar M\to G\times Y$
is some smooth mapping into $G\times Y$. Under the mappings scheme (4) the
expression (3) takes the following form:
\begin{equation}
\left.s^*\Omega^j\right|_{\bar M}:=
\left.\mu^*\bar\Omega^j\right|_{\bar M}:\Rightarrow 0
\end{equation}
for all $j=\overline{1,r}$ upon the integral submanifold $\bar M\subset M$,
where $\Omega^j\in \Lambda^2(M),\,j=\overline{1,r},$ is some {\it a priori}
given system of 2-forms on $M$.

Assume further that a set $\{\alpha_j\in\Lambda^2(M):\,
j=\overline{1,m_\alpha}\}$ is a basis of two-forms $\{\Omega^j\in
\Lambda^2(M): \,j=\overline{1,r}\}$, generating the ideal
${\cal I}(\alpha)\subset\Lambda(M)$. The ideal ${\cal I}(\alpha)$ should
be completely integrable within the Cartan criterium, because due to the
set of equations $d\Omega^j\in{\cal I}(\Omega), \,j=\overline{1,r}$,
following from (3), giving ${\cal I}(\Omega)\equiv 0$ on $\bar M$,
from the scheme (4) it follows that $d{\cal I}(\alpha)\subset
{\cal I}(\alpha)$ since $s^*{\cal I}(\alpha)=\mu^*{\cal I}(\bar\Omega)$.

To def\/ine now   a criterium for  a Lie group action $G\times Y
\stackrel{\rho}{\to} Y$ to
generate a representation of the Lie group $G$, we need to build
the ideal ${\cal I}(\alpha, \beta)\subset\Lambda(M\times Y)$, corresponding
to (2) and (5), for a some set of forms $\beta^j\in\Lambda^1(M\times Y), \,
j=\overline{1,n}$, where $s^*\beta_j:=\mu^*\bar\beta_j\in
\Lambda(\bar M\times Y), \,j=\overline{1,n}$, and to insist it to be closed in
$\Lambda(M\times Y)$, that is $d{\cal I}(\alpha, \beta)\subset
{\cal I}(\alpha, \beta)$, or
\begin{equation}
d\beta^j=\sum^{m_\alpha}_{k=1}f_k^j\alpha^k+\sum^n_{i=1}g^j_i\land \beta^i
\end{equation}
for all $j=\overline{1,n}$ and some $f^j_k\in\Lambda^0(M\times Y), \,
k=\overline{1,m_\alpha}, \,\,g^j_i\in\Lambda^1(M\times Y), \,
i,j=\overline{1,n}$. The condition (6) assures that there exist some smooth
submanifold $\bar M(Y)\subset M\times Y$, on which a nonlinear Lie group $G$
representation acts exactly. Thereby we have stated that the following
theorem is valid [4].

\medskip

\noindent
{\bf Theorem 2.} {\it The system $\{\beta\}$ of Cartan's one-forms
$\beta^j\in\Lambda^1(M\times Y), \,j=\overline{1,n}$, generated by the
mapping scheme (4), describes an exact nonlinear Lie group $G$ representation
on a manifold $Y$ if and only if the adjoint ideal
${\cal I}(\alpha, \beta)$ generated by the system $\{\beta\}$ and a basic system
$\{\alpha\}$ of the ``curvature'' 2-forms  $\Omega^j\in \Lambda^2(M), \,
j=\overline{1,r}$, of (5), is closed together with the corresponding ideal
${\cal I}(\alpha)={\cal I}(\alpha,0)$ in the  Grassmann algebras
$\Lambda(M\times Y)$ and $\Lambda(M)$ correspondingly.}

\medskip

Going out of the results stated above, it is naturally to make some specialization
of Cartan's geometric construction by means of the  theory of principal
f\/iber bundles [5]. To proceed with, let us try to interpret the  Cartan
dif\/ferential system $\{\beta\}$ on $M\times Y$ as one generating a linear
$(r\times r)$ - matrix adjoint representation [6,10]
of the Lie algebra $\cal G$, putting the functions
$\xi^j_i(y):=\sum\limits^r_{k=1}c^j_{ik}y^k, \,i,j=\overline{1,r},$ when
$\mbox{dim}\,Y=n:=r:$
\begin{equation}
\beta^j:\Rightarrow dy^j+\sum^r_{i,k=1}c^j_{ik}y^kb^i(z)\in
\Lambda^1(M\times Y),
\end{equation}
where $z\in M$, and 1-forms $b^i(z)$ on $M$ satisfy the necessary embedding
conditions \linebreak $s^*b^i=\mu^*\bar\omega^i$ upon $\bar M\subset M$ for all
$i,j=\overline{1,r}$ in accordance with the scheme (4).

The Lie group $G$ acts on the linear $r$-dimensional space $Y$ by the usual
left shifts as follows: $Y\times G\ni y\times a\stackrel{\rho}{\to}ay\in Y$
for all $a\in G$. Whence we can easily deduce the following inf\/initesimal
shifts in the Lie group $G$:
\[
da^j_k+\sum^r_{s=1}c^j_{si}b^s(z)a^i_k\in \Lambda^1(M\times G).
\]
These expressions ultimately engender the next ${\cal G}$-valued
$Ad$-invariant 1-form
$\omega$ on $M\times G$ via the isomorphic mapping
$\rho^*: \Lambda^1(M\times Y)\to \Lambda^1(M\times G)\otimes\cal G$:
\begin{equation}
\{\beta\}:\stackrel{\rho^*}{\longrightarrow}\omega:=a^{-1}da+Ad_{a^{-1}}\Gamma(z),
\end{equation}
where the one-forms matrix $\Gamma(z):=\|\Gamma^j_k(z)\|, \,
j,k=\overline{1,r}$, belongs to the $(r\times r)$-matrix representation of the
Lie algebra $\cal G$ due to  construction:
$\|\Gamma^j_k(z)\|:=\|\sum\limits^r_{i=1}c^j_{ik}b^i(z)\|\in T^*(M)\otimes\cal G$.
The results above one can naturally interpret as a way of def\/ining
[5, 7]
some $\cal G$-valued connection $\Gamma$ upon a principal f\/ibered space
$P(M;G)$, carrying the $\cal G$-valued connection 1-form (8). The
corresponding Cartan's 1-forms determine the horizontal subspace of the
parallel transporting vectors of the  f\/iber bundle  $P(M;G,Y)$
associated with $P(M;G)$
 according to the general theory [5] of f\/ibered spaces with connections.

Thus, we have built the $\cal G$-valued connection 1-form (8) at a point
$(z,a)\in P(M;G)$ as $\omega:=\bar\omega(a)+Ad_{a^{-1}}\Gamma(z)$, where
$\bar\omega(a)\in T^*(G)\otimes{\cal G}$ is the standard Maurer-Cartan
left- invariant ${\cal G}$- valued 1-form on the Lie group $G$. The
connection 1-form (8) is vanishing upon the above mentioned horizontal
subspace, consisting of vector f\/ields on $P(M;G)$, which generate
a Lie group $G$ representation on the space
$Y$ . This means, that this horizontal
subspace necessarily def\/ines a completely integrable dif\/ferential system
on $P(M;G)$, or equivalently, the corresponding curvature $\Omega\in
\Lambda^2(M)\otimes{\cal G}$ of the connection $\Gamma$ is vanishing upon
the integral submanifold $\bar M\subset M$:
\begin{equation}
\Omega:=d\omega+\omega\land\omega=
Ad_{a^{-1}}(d\Gamma(z)+\Gamma(z)\land\Gamma(z))=\frac12 Ad_{a^{-1}}\left.
\sum^m_{j=1}\Omega_{jk}dz^j\land dz^k\right|_{\bar M}\Rightarrow 0.
\end{equation}
from where we obtain
\begin{equation}
\ba{l}
\ds \Omega_{ij}(z):=\frac{\partial\Gamma_j(z)}{\partial z_i} -
\frac{\partial\Gamma_i(z)}{\partial z_j}+[\Gamma_i(z),\Gamma_j(z)],\\[4mm]
\ds
\Gamma(z):=\sum\limits^m_{j=1}\Gamma_j(z)dz^j:=\sum\limits^m_{j=1}
\sum\limits^r_{k=1} \Gamma^k_j(z)dz^jA_k .
\ea
\ee
The vanishing curvature $\Omega$ (9) upon the submanifold $\bar M\subset M$
is easily explained by means of the followimg commuting
diagram:
\begin{equation}
\begin{array}{c@{}c@{}c@{}c@{}c@{}c@{}c}
T^*(G) & \stackrel{\mu^*}{\longrightarrow} & T^*(P(\bar M;G)) &
\stackrel{s^*}{\longleftarrow} & T^*(P(M;G))  &
\stackrel{\rho^*}{\longleftarrow} & T^*(P(M;G,Y))
\\
\downarrow &  & \downarrow &  & \downarrow &  & \downarrow \\
G & \stackrel{\mu}{\longleftarrow} & P(\bar M;G) &
\stackrel{s}{\longrightarrow} & P(M;G) & \stackrel{\rho}{\longrightarrow} &
P(M;G,Y)
\end{array}
\end{equation}
We can now derive from (12), that due to (8)
\begin{equation}
\rho^*\{\beta\}=\omega, \quad s^*\Gamma^j_k=\sum^r_{i=1}c^j_{ik}\mu^*\bar
\omega^i\,\,\Rightarrow\,\,s^*\Omega=\mu^*\bar\Omega=0,
\end{equation}
giving rise to the implication (9) upon $\bar M$.

Thus, if some integrable ideal ${\cal I}(\alpha)\subset\Lambda(M)$ is
{\it a priori} given
 on the manifold $M$, we can take the corresponding to (9)
equation in $\Lambda(M)$:
\[
\sum^m_{j,k=1}\Omega_{jk}dz^j\land dz^k\subset{\cal I}(\alpha)
\otimes {\cal G}
\]
both as determining the ${\cal G}$-valued 1-forms $\Gamma_j(z)\in
T^*(M)\otimes{\cal G}, \,j=\overline{1,m}$, and as determining a Lie algebra
structure of ${\cal G}$, taking into account the holonomy Lie group reduction
theorem of Ambrose, Singer and Loos [9, 10]. Namely, the holonomy Lie algebra
${\cal G}(h)\subset {\cal G}$ being generated by covariant derivatives
composition of the ${\cal G}$-valued curvature form
$\Omega\in T^*(M)\otimes{\cal G}$:
\begin{equation}
{\cal G}(h):= span_{\bf R}\{\nabla^{j_1}_1 \nabla^{j_2}_2\ldots
\nabla^{j_n}_n\Omega_{si}\in{\cal G}: \,\,j_k\in{\bf Z_+}, \,
s,i,k=\overline{1,n}\}
\end{equation}
where, by def\/inition, the covariant derivative $\nabla_j:\Lambda(M)\to
\Lambda(M), \,j=\overline{1,n}$, is given as follows
\begin{equation}
\nabla_j:=\partial/\partial z^j+\Gamma_j(z).
\end{equation}
If the identity ${\cal G}(h)\equiv span_{\bf R}\{\Omega_{sl}\in{\cal G}: \,\,
s,l=\overline{1,n}\}$ takes place, that is the inclusion
$[{\cal G}(h),{\cal G}(h)]\subset{\cal G}(h)$ is reached, the holonomy Lie
algebra ${\cal G}(h)$ is called perfect. Thus, we can formulate the following
equivalence theorem.

\medskip

\noindent
{\bf Theorem 3}. {\it Given a closed  ideal ${\cal I}(\alpha)$ on
a manifold $M$, $d{\cal I}(\alpha)\subset{\cal I}(\alpha)$, its 1-forms
augmentation ${\cal I}(\alpha,\beta)$ on $M\times Y$ by means of a special
set $\{\beta\}$ of 1-forms
\begin{equation}
\{\beta\}:=\left\{\beta^j=dy^j+\sum^n_{k=1}\xi^j_k(y)b^k(z):\,
b^j(z)\in T^*(M), \,j=\overline{1,n}\right\},
\end{equation}
compatible with the scheme (11), is integrable within Frobenius-Cartan
criterium if and only if there exists some Lie group $G$ action on $Y$, such
that the adjoint connection (8) on a f\/ibered space $P(M;G)$ with the structure
group $G$ is vanishing upon the integral submanifold $\bar M\subset M$ of the
ideal ${\cal I}(\alpha)\subset\Lambda(M)$. The latter can serve as the
algorithm of determining the structure of the Lie group $G$ basing
on the holonomy Lie algebra reduction theorem of Ambrose-Singer-Loos
[9, 10].}

\medskip

If the conditions of Theorem 3 are fulf\/illed, the set of 1-forms $\{\beta\}$
(15) generates a representation of the
Lie group $G$ upon the analytical manifold $Y$ according to the Cartan theorem 1.
The Lie algebra ${\cal G}$ of
the Lie group $G$ can be reduced to the holonomy Lie algebra ${\cal G}(h)$,
generated via (13) by the curvature 2-form $\Omega$ of the connection $\Gamma$
on the principal f\/iber bundle $P(M;G)$ built above.

\section{An ef\/fective Maurer-Cartan one-forms construction}

To proceed further in study of the integrability of Lie-invariant geometric
objects generated by the scheme (4) with some mapping $s: \bar M\to G$, one
needs to have an ef\/fective way of construction corresponding to the Lie
algebra ${\cal G}\simeq T^*_e(G)$ the Maurer-Cartan forms $\bar\omega^j(a;da)
\in T^*_a(G)\otimes{\cal G},\,
j=\overline{1,r}$. Below we will describe an ef\/fective direct
procedure of building these forms on $G$.

Let be given a Lie group $G$ with the Lie algebra ${\cal G}\simeq T_e(G)$,
whose basis is a set $\{A_i\in{\cal G}:\, i=\overline{1,r}\}$, where
$r=\mbox{dim}\,G\equiv \mbox{dim}\,{\cal G}$. Let also a set
$U_0\subset\{a^i\in{\bf R}:\,i=\overline{1,r}\}$ be some open neighborhood of the zero
point in ${\bf R}^r$. The exponential mapping $\exp:\,U_0\to G_0$, where
by def\/inition
\begin{equation}
{\bf R}^r\supset U_0\ni (a^1, \ldots , a^r):\stackrel{\exp}{\longrightarrow}
\exp\left(\sum^r_{i=1}a^iA_i\right):=a\in G_0\subset G,
\end{equation}
is an analytical mapping of the whole $U_0$ on some open neighborhood $G_0$
of the unity element $e\in G$. From (16) it is easy to f\/ind that
$T_e(G)=T_e(G_0)\simeq {\cal G}$, where $e:=\exp(0)\in G$. Def\/ine now the
following left invariant ${\cal G}$- valued dif\/ferential one-form on
$G_0\subset G$:
\begin{equation}
\bar\omega(a;da):=a^{-1}da=\sum^r_{j=1}\bar\omega^j(a,da)A_j\in{\cal G},
\end{equation}
where $\bar\omega^j(a;da)\in T^*_a(G), \,a\in G_0, \,j=\overline{1,r}
$. To build ef\/fectively
the unknown forms $\{\bar\omega^j(a;da):\, j=\overline{1,r}\}$, let us
consider the following analytical one-parameter one-form
$\bar\omega_t(a;da):=\bar\omega(a_t;da_t)$ on $G_0$, where
$a_t;=\exp\left(t\sum\limits^r_{i=1}a^iA_i\right),\,  t\in[0,1]$, and
dif\/ferentiate this form with respect to the parameter $t\in[0,1]$. We will
get that
\begin{equation}
\ba{l}
\ds d\bar\omega_t/dt=-\sum\limits^r_{j=1}a^jA_j a^{-1}_tda_t+
\sum\limits^r_{j=1}a^{-1}_ta_tda^jA_j+
\sum\limits^r_{j=1}a^{-1}_tda_ta^jA_j\\[3mm]
\ds \phantom{d\bar\omega_t/dt}=-\sum\limits^r_{j=1}a^j[A_j,\bar\omega_t]+
\sum\limits^r_{j=1}A_jda_j.
\ea
\ee
Having used the Lie identity $[A_j, A_k]=\sum\limits^r_{i=1}c^i_{jk}A_i, \,
j,k=\overline{1,r}$, and the right hand side of (17) in form
\begin{equation}
\bar\omega^j(a;da):=\sum^r_{k=1}\bar\omega^j_k(a)da^k,
\end{equation}
we ultimately obtain that
\begin{equation}
\frac d{dt}(t\bar\omega^j_i(ta))=\sum^r_{k=1}{\cal A}^j_k
t\bar\omega^k_i(ta)+\delta^j_i,
\end{equation}
where the matrix ${\cal A}^k_i, \,i,k=\overline{1,r}$, is def\/ined as follows:
\begin{equation}
{\cal A}^k_i:=\sum^r_{j=1}c^k_{ij}a^j.
\end{equation}
Thus, the matrix $W^j_i(t):=t\bar\omega^j_i(ta), \,i,j=\overline{1,r}$,
satisf\/ies the following from (20) dif\/ferential equation [6]
\begin{equation}
dW/dt={\cal A}W+E, \quad \left.W\right|_{t=0}=0,
\end{equation}
where $E=\|\delta^j_i\|$ is the unity matrix. The solution of (22)
is representable as
\begin{equation}
W(t)=\sum^\infty_{n=1}\frac{t^n}{n!}{\cal A}^{n-1}
\end{equation}
for all $t\in[0,1]$. Whence, recalling the above def\/inition of the matrix
$W(t)$, we obtain easily that
\begin{equation}
\left.\bar\omega^j_k(a)=W^j_k(t)\right|_{t=1}=\sum^\infty_{n=1}
(n!)^{-1}{\cal A}^{n-1}.
\end{equation}
Thereby the task of f\/inding the Maurer-Cartan one-form for a given Lie
algebra ${\cal G}$ is solved in the ef\/fective and constructive way, being
at the same time completely algebraic.

Therefore, the following theorem solves the problem of f\/inding in an
ef\/fective algebraic way corresponding to a Lie algebra ${\cal G}$ the left
invariant one-form $\bar\omega(a;da)\in T^*_a(G)\otimes{\cal G}$ at any
$a\in G$:

\medskip

\noindent
{\bf Theorem 4}.  {\it Let's be given a Lie algebra ${\cal G}$ with the structure
constants $c^k_{ij}\in{\bf R}$, $i,j,k=\overline{1,r=\mbox{dim}\,{\cal G}}$, related
to some basis $\{A_j\in{\cal G}:\,j=\overline{1,r}\}$. Then the adjoint
to ${\cal G}$ left-invariant Maurer-Cartan one-form $\bar\omega(a;da)$
is built as follows:
\begin{equation}
\bar\omega(a;da)=\sum^r_{k,j=1}A_j\bar\omega^j_k(a)da^k,
\end{equation}
where the matrix $W:=\|\bar w^j_k(a)\|, \,j,k=\overline{1,r}$,   is given
exactly as
\begin{equation}
W=\sum^\infty_{n=1}(n!)^{-1}{\cal A}^{n-1}, \quad
{\cal A}^j_k:=\sum^r_{i=1}c^j_{ki}a^i.
\end{equation}}

\medskip

Below we shall try to use the experience gained above in solving an
analogous problem of the theory of connections over a principal f\/iber bundle
$P(M;G)$ as well as over associated with it a f\/iber bundle $P(M;Y,G)$.

\section{General structure of integrable one-forms augmenting the
two-forms associated with a closed ideal in the Grassmann algebra}

Given  two-forms generating a closed ideal ${\cal I}(\alpha)$ in the Grassmann
algebra $\Lambda(M)$, we will denote as above by ${\cal I}(\alpha,\beta)$
an augmented ideal in $\Lambda(M;Y)$, where the manifold $Y$ will be called
in further the representation space of some adjoint Lie group $G$ action:
$G\times Y\stackrel{\rho}{\to}Y$.  We can f\/ind therefore the determining
relationships for the set of one-forms $\{\beta\}$ and 2-forms $\{\alpha\}$
\be
\ba{l}
\{\alpha\}:=\{\alpha^j\in\Lambda^2(M):\,j=\overline{1,m_\alpha}\},
\\[2mm]
\{\beta\}:=\{\beta^j\in\Lambda^1(M\times
Y):\,j=\overline{1,n=\mbox{dim}\,Y}\},
\ea
\ee
satisfying such equations:
\be
\ba{l}
\ds d\alpha^i=\sum\limits^{m_\alpha}_{k=1}a^i_k(\alpha)\land\alpha^k,
\\[3mm]
\ds d\beta^j=\sum^{m_\alpha}_{k=1}f^j_k\alpha^k+
\sum\limits^n_{s=1}\omega^j_s\land\beta^s,
\ea
\ee
where $a^i_k(\alpha)\in\Lambda^1(M), \,f^j_k\in\Lambda^0(M\times Y)$ and
$\omega^j_s\in\Lambda^1(M\times Y)$ for all $i,k=\overline{1,m_\alpha}, \,
j,s=\overline{1,n}$. Since the identity $d^2\beta^j\equiv 0$ takes place
for all $j=\overline{1,n}$, from (28) we deduce the following relationship:
\begin{equation}
\sum^n_{k=1}\left(d\omega^j_k+\sum^n_{s=1}\omega^j_s\land\omega^s_k\right)
\land\beta^k+
\sum^{m_\alpha}_{s=1}\left(df^j_s+
\sum^n_{k=1}\omega^j_kf^k_s+\sum^{m_\alpha}_{l=1}f^j_la^l_s(\alpha)
\right)\land\alpha^s\equiv 0.
\ee
As a result of (29) we obtain that
\be
\ba{l}
\ds d\omega^j_k+\sum\limits^n_{s=1}\omega^j_s\land\omega^s_k\in{\cal I}(\alpha,\beta),
\\[3mm]
\ds df^j_s+\sum\limits^n_{k=1}\omega^j_kf^k_s+
\sum\limits^{m_\alpha}_{l=1} f^j_la^l_s(\alpha)
\in{\cal I}(\alpha, \beta)
\ea
\ee
for all $j,k=\overline{1,n}, \, s=\overline{1,m_\alpha}$. The second inclusion
in (30) gives a possibility to def\/ine the 1-forms
$\theta^j_s:=\sum\limits^{m_\alpha}_{l=1}f^j_la^l_s(\alpha)$ satisfying the next
inclusion:
\begin{equation}
d\theta^j_s+\sum^n_{k=1}\omega^j_k\land\theta^k_s\in{\cal I}(\alpha, \beta)
\oplus\sum^{m_\alpha}_{l=1}f^j_lc^l_s(\alpha),
\end{equation}
which we obtained having used the identities $d^2\alpha^j\equiv 0, \,
j=\overline{1,m_\alpha}$, in the form $\sum\limits^m_{s=1}c^j_s(\alpha)\land\alpha^s
\equiv 0$,
\begin{equation}
c^j_s(\alpha)=da^j_s(\alpha)+\sum^{m_\alpha}_{k=1}
a^j_l(\alpha)\land a^l_s(\alpha),
\end{equation}
following from (28). Let us suppose further that as $s=s_0$ the 2-forms
$c^j_{s_0}(\alpha)\equiv0$ for all $j=\overline{1,m_\alpha}$. Then as $s=s_0$,
we can def\/ine a set of 1-forms $\theta^j:=\theta^j_{s_0}\in
\Lambda^1(M\times Y), \,j=\overline{1,n}$, satisfying the exact inclusions:
\begin{equation}
d\theta^j+\sum^n_{k=1}\omega^j_k\land\theta^k:=\Theta^j\in{\cal I}(\alpha,\beta)
\end{equation}
together with a set of inclusions for 1-forms $\omega^j_k\in\Lambda^1(M\times Y)$
\begin{equation}
d\omega^j_k+\sum^n_{s=1}\omega^j_s\land\omega^s_k:=\Omega^j_k\in
{\cal I}(\alpha,\beta)
\end{equation}
As it follows from the general theory [5] of connections on the f\/ibered
frame space $P(M;GL(n))$ over a base manifold $M$, we can interpret the
equations (34) as the equations def\/ining the curvature 2-forms $\Omega^j_k
\in\Lambda^2(P)$, as well as interpret the equations (33) as those, def\/ining
the torsion 2-forms $\Theta^j\in\Lambda^2(P)$. Since ${\cal I}(\alpha)=0=
{\cal I}(\alpha,\beta)$ upon the integral submanifold $\bar M\subset M$, the
reduced f\/ibered frame space $P(\bar M;GL(n))$ will have the f\/lat curvature
and be torsion free, being as a result, completely trivialized on
$\bar M\subset M$. Consequently, we can formulate the following theorem.

\medskip

\noindent
{\bf Theorem 5.} {\it Let the condition above on the ideals ${\cal I}(\alpha)$
and ${\cal I}(\alpha,\beta)$ be fulf\/illed. Then the set of 1-forms $\{\beta\}$
generates the integrable augmented ideal ${\cal I}(\alpha,\beta)\subset
\Lambda(M\times Y)$ if and only if there exists some curvature 1-form
$\omega\in\Lambda^1(P)\otimes{\cal G}l(n)$ and torsion 1-form $\theta\in
\Lambda^1(P)\otimes{\bf R}^n$ on the adjoint f\/ibered frame space $P(M;GL(n))$,
satisfying the inclusions
\be
\ba{l}
d\omega+\omega\land\omega\in{\cal I}(\alpha,\beta)\otimes{\cal G}l(n),
\\[2mm]
d\theta+\omega\land\theta\in{\cal I}(\alpha,\beta)\otimes{\bf R}^n.
\ea
\ee
Upon the reduced f\/ibered frame space $P(\bar M;GL(n))$ the
corresponding curvature and torsion are vanishing, where $\bar
M\subset M$ is the integral submanifold of the ideal ${\cal
I}(\alpha)\subset\Lambda(M)$. }

\medskip

We can see from Theorem 5 that some its conditions coincide with
those of Theorem 3, concerning the properties of adjoint curvature forms
$\omega\in\Lambda^1(P)\otimes{\cal G}$. Thus, the condition of existing
some curvature 1-form $\omega\in\Lambda^1(P)\otimes{\cal G}$, whose
curvature form $\Omega\in\Lambda^2(P)\otimes{\cal G}$ must necessarily
vanish upon the integral submanifold of the ideal ${\cal I}(\alpha)\subset
\Lambda(M)$. The nature of the  second inclusion of (35) is at present not
completely understood, namely the condition of existence of the integrable
augmented ideal ${\cal I}(\alpha,\beta)\subset\Lambda(M\times Y)$. This
problem is under started view of an article under preparation. Below we will
analyse in detail some special examples [7, 8] of the construction
suggested above,
concerned with the integrable dynamical systems, given on some invariant
jet-submanifolds.

\section{The Cartan's invariant geometric object structure of Lax integrable
nonlinear dynamical systems in partial derivatives}

Consider at the beginning some set $\{\beta\}$ def\/ining a
Cartan's Lie group $G$ invariant object
on a manifold $M\times Y$:
\begin{equation}
\beta^j:=dy^j+\sum^r_{k=1}\xi^j_k(y)b^k(z),
\end{equation}
where $i=\overline{1,n=\mbox{dim}\,Y}, \,r=\mbox{dim}\,G$,
satisfying the mapping scheme (4)
with a chosen integral submanifold $\bar M\subset M$. This means, that the set
(36) def\/ines on the manifold $Y$ a set $\{\xi\}$ of vector f\/ields, compiling a
representation $\rho:{\cal G}\to\{\xi\}$ of a given Lie algebra $\cal G$,
that is vector f\/ields
$\ds \xi_s:=\sum\limits^n_{j=1}\xi^j_s(y)\frac\partial{\partial y^j}\in\{\xi\},
\,s=\overline{1,r}$, enjoy the following Lie algebra $\cal G$ relationships
\begin{equation}
[\xi_s,\xi_l]=\sum^r_{k=1}c^k_{sl}\xi_k
\end{equation}
for all $s,l,k=\overline{1,r}$. We can now compute the dif\/ferentials
$d\beta^j\in\Lambda^2(M\times Y), \, j=\overline{1,n}$, using (36) and (37)
as follows:
\begin{equation}
\ba{l}
\ds d\beta^j=\sum\limits^n_{l=1}\sum\limits^r_{k=1}
\frac{\partial\xi^j_k(y)}{\partial y^l}
\left(\beta^l-\sum^r_{s=1}\xi^l_s(y)b^s(z)\right)\land b^k(z)+
\sum\limits^n_{l=1}\sum\limits^r_{k=1}\xi^j_k(y)db^k(z)\\[5mm]
\ds = \sum\limits^n_{l=1}\sum\limits^r_{k=1}
\frac{\partial\xi^j_k(y)}{\partial y^l} \beta^l\land b^k(z)-\!
\sum\limits^n_{l=1}\sum\limits^r_{k,s=1}
\frac{\partial\xi^j_k(y)}{\partial y^l}
\xi^l_s(y)b^s(z)\land b^k(z)+\sum\limits^r_{k=1}\xi^j_k(y)db^k(z)\!
\\[5mm]
\ds =\sum\limits^n_{l=1}\sum\limits^r_{k=1}\frac{\partial
  \xi^j_k(y)}{\partial y^l}
\beta^l\land b_k(z)+\frac12\sum\limits^n_{l=1}\sum\limits^r_{k,s=1}
\left[\frac{\partial  \xi^j_k(y)}{\partial y^l}\xi^l_s(y)-
\frac{\partial  \xi^j_s(y)}{\partial y^l}\xi^l_k(y)\right]
\\[5mm]
\ds \times db^k(z)\land db^s(z)+\sum\limits^r_{k=1}\xi^j_k(y)db^k(z)
\ea
\ee
$\Rightarrow$
\[
\ba{l}
\ds
\sum\limits^n_{l=1}\sum\limits^r_{k=1}\frac{\partial
\xi^j_k(y)}{\partial y^l}
\beta^l\land
b_k(z)+\frac12\sum\limits^r_{k,s=1}[\xi_s,\xi_k]^jdb^k(z)\land
db^s(z) \\[5mm]
\ds +\sum\limits^r_{k=1}\xi^j_k(y)db^k(z)\Rightarrow
 \sum\limits^n_{l=1}\sum\limits^r_{k=1}
\frac{\partial  \xi^j_k(y)}{\partial y^l}
\beta^l\land b_k(z)\\[5mm]
\ds +\frac12\sum\limits^n_{l=1}\sum\limits^r_{k,s=1}
c^l_{ks}\xi^j_ldb^k(z)\land db^s(z)+
\sum\limits^r_{k=1}\xi^j_k(y)db^k(z)\Rightarrow
\sum\limits^n_{l=1}\sum\limits^r_{k=1}\frac{\partial \xi^j_k(y)}{\partial y^l}
\beta^l\land b_k(z)\\[5mm]
\ds +\sum\limits^r_{l=1}\xi^j_l\left(db^l(z)+\frac12\sum\limits^r_{k,s=1}
c^l_{ks}db^k(z)\land db^s(z)\right):\in{\cal I}(\alpha,\beta)\subset
\Lambda(M\times Y),
\ea
\]
where $\{\alpha\}\subset\Lambda^2(M)$ is some {\it a priori} given integrable
system of 2-forms on $M$,  vanishing upon the integral submanifold
$\bar M\subset M$. It is obvious that inclusions (38) take place if and
only if the following conditions are fulf\/illed: for all $j=\overline{1,r}$
\begin{equation}
db^j(z)+\frac12\sum^r_{k,s=1}c^j_{ks}db^k(z)\land db^s(z)\in{\cal I}(\alpha).
\end{equation}
The inclusions (39) mean in particular, that upon the integral submanifold
$\bar M\subset M$ of the ideal ${\cal I}(\alpha)\subset\Lambda(M)$ the
equalities
\begin{equation}
\mu^*\bar \omega^j\equiv s^*b^j,
\end{equation}
are true,
where $\bar \omega^j\in T^*_e(G), \,j=\overline{1,r}$, are the left invariant
Maurer-Cartan forms on the invariance Lie group $G$. Thus, due to inclusions
(39) all conditions of Cartan's Theorem 1 are enjoyed, giving rise to a
possibility to obtain the set of forms $b^j(z)\in\Lambda^1(M)$ in an explicit
form. To do this, let us def\/ine a $\cal G$-valued curvature 1-form
$\omega\in \Lambda^1(P(M;G))\otimes\cal G$ as follows
\begin{equation}
\omega:=Ad_{a^{-1}}\left(\sum^r_{j=1}A_jb^j\right)+\bar\omega
\end{equation}
where $\bar\omega\in\cal G$ is the standard Maurer-Cartan 1-form on $G$,
built in Chapter 2. This
1-form satisf\/ies followed by (39) the canonical structure inclusion (9) for
$\Gamma:=\sum\limits^r_{j=1}A_jb^j\in\Lambda^1(M)\otimes\cal G$:
\begin{equation}
d\Gamma+\Gamma\land\Gamma\in {\cal I}(\alpha)\otimes{\cal G},
\end{equation}
serving as a main relationships determining the form (41) in accordance
with results of Chapter 3. To proceed further we need to give the
set of 2-forms $\{\alpha\}\subset \Lambda^2(M)$ in explicit form.

\medskip

\noindent
{\bf Example 1. The Burgers dynamical system.}

Let's be given the following Burgers dynamical system on a functional manifold
\linebreak $M\subset C^\infty({\bf R};{\bf R})$:
\begin{equation}
u_t=uu_{x}+u_{xx},
\end{equation}
where $u\in M,\,   t\in{\bf R}$ is an evolution parameter. The f\/low (43) on $M$ can
be recast into a set of 2-forms $\{\alpha\}\subset\Lambda^2(J({\bf R}^2;{\bf R}))$
upon the adjoint jet-manifold $J({\bf R}^2;{\bf R})$ as follows:
\begin{equation}
\ba{l}
\{\alpha\}=\left\{du^{(0)}\land dt-u^{(1)}dx\land dt=\alpha^1, \,\,
du^{(0)}\land dx+u^{(0)}du^{(0)}\land dt
\right.\\[2mm]
\phantom{\{\alpha\}=} \left.
+du^{(1)}\land dt=\alpha^2:\,\,
\left(x,t;u^{(0)},u^{(1)}\right)^\tau\in M^4\subset J^1({\bf R}^2;{\bf R})
\right\},
\ea
\ee
where $M^4$ is some f\/inite-dimensional submanifold in $J^1({\bf R}^2;{\bf R}))$ with
coordinates  $\Bigl(x, t$, $u^{(0)}=u, u^{(1)}=u_x\Bigr)$.
The set of 2-forms
(44) generates the closed ideal ${\cal I}(\alpha)$, since
\begin{equation}
d\alpha^1=dx\land\alpha^2-u^{(0)}dx\land\alpha^1, \quad d\alpha^2=0,
\end{equation}
the integral submanifold $\bar M=\{x,t\in{\bf R}\}\subset M^4$ being def\/ined by
the condition ${\cal I}(\alpha)=0$. We now look for a reduced "curvature"
1-form $\Gamma\in\Lambda^1(M^4)\otimes\cal G$, belonging to some not yet
determined Lie algebra $\cal G$. This 1-form can be represented using
(44), as follows:
\begin{equation}
\Gamma:=b^{(x)}(u^{(0)}, u^{(1)})dx+b^{(t)}(u^{(0)},u^{(1)})dt,
\end{equation}
where elements $b^{(x)}, b^{(t)}\in\cal G$ satisfy such determining equations,
engendered by (42):
\begin{equation}
\ba{l}
\ds \frac{\partial b^{(x)}}{\partial u^{(0)}}du^{(0)}\land dx+
\frac{\partial b^{(x)}}{\partial u^{(1)}}du^{(1)}\land dx+
\frac{\partial b^{(t)}}{\partial u^{(0)}}du^{(0)}\land dt
\\[4mm]
\qquad \ds +\frac{\partial b^{(t)}}{\partial u^{(1)}}du^{(1)}\land dt+
[b^{(x)},b^{(t)}]dx\land dt\equiv\Omega\\[3mm]
\ds \Rightarrow \quad  g_1(du^{(0)}\land dt-u^{(1)}dx\land dt)+
g_2(du^{(0)}\land dx\\[2mm]
\phantom{\Rightarrow}
\ds \quad +u^{(0)}du^{(0)}\land dt+du^{(1)}\land dt)
\in{\cal I}(\alpha)\otimes{\cal G}
\ea
\ee
for some $\cal G$-valued functions $g_1, \,g_2$ on $M$. From (47)
it follows that
\begin{equation}
\ba{l}
\ds \frac{\partial b^{(x)}}{\partial u^{(0)}}=g_2, \quad
\frac{\partial b^{(x)}}{\partial u^{(1)}}=0, \quad
\frac{\partial b^{(t)}}{\partial u^{(0)}}=g_1+g_2u^{(0)},\\[4mm]
\ds \frac{\partial b^{(t)}}{\partial u^{(1)}}=g_2, \quad
[b^{(x)},b^{(t)}]=-u^{(1)}g_1.
\ea
\ee
The set (48) has the following unique solution
\begin{equation}
\ba{l}
\ds b^{(x)}=A_0+A_1u^{(0)},\\[3mm]
\ds b^{(t)}=u^{(1)}A_1+\frac{{u^{(0)}}^2}2A_1+[A_1,A_0]u^{(0)}+A_2,
\ea
\ee
where $A_j\in{\cal G}, \, j=\overline{0,2}$, are some constant elements on $M$
of a Lie algebra $\cal G$ under search, enjoying the next Lie structure
equations:
\be
\ba{l}
[A_0,A_2]=0,\\[2mm]
[A_0,[A_1,A_0]]+[A_1,A_2]=0,\\[2mm]
\ds [A_1,[A_1,A_0]]+\frac12[A_0,A_1]=0.
\ea
\ee
From (48) one can see that the curvature 2-form $\Omega \in span_{\bf R}
\{A_1,[A_0,A_1]:\, A_j\in{\cal G},\, j=\overline{0,1}\}$. Therefore,
reducing via the Ambrose-Singer theorem the associated principal f\/ibered
frame space $P(M;G=GL(n))$ to the principal f\/iber bundle $P(M;G(h))$,
where $G(h)\subset G$ is the corresponding holonomy Lie group of the
connection $\Gamma$ on $P$, we need to satisfy the following conditions
for the set ${\cal G}(h)\subset\cal G$ to be a Lie subalgebra in ${\cal G}:\,
\nabla_x^m\nabla_t^n\Omega\in{\cal G}(h)$ for all $m,n\in{\bf Z}_+$.

Let us try now to close the above transf\/initive procedure requiring that
\begin{equation}
{\cal G}(h)={\cal G}(h)_0:=span_{\bf R}
\{\nabla^m_x\nabla^n_x\Omega\in{\cal G}: \,m+n=0\}
\end{equation}
This means that
\begin{equation}
{\cal G}(h)_0=span_{\bf R}\{A_1, A_3=[A_0,A_1]\}.
\end{equation}
To enjoy the set of relations (50) we need to use expansions over the
basis (52) of the external elements $A_0,A_2\in{\cal G}(h)$:
\be
A_0=q_{01}A_1+q_{13}A_3,  \qquad A_2=q_{21}A_1+q_{23}A_3.
\ee
Substituting expansions (53) into (50), we get that $q_{01}=q_{23}=\lambda, \,\,
q_{21}=-\lambda^2/2$ and $q_{03}=-2$ for some  arbitrary real parameter
$\lambda\in{\bf R}$, that is ${\cal G}(h)=span_{\bf R}\{A_1,A_3\}$, where
\begin{equation}
[A_1,A_3]=A_3/2; \qquad A_0=\lambda A_1-2A_3,
\qquad A_2=-\lambda^2A_1/2+\lambda A_3.
\end{equation}
As a result of (54) we can state that the holonomy Lie algebra ${\cal G}(h)$
is a real two-dimensional one, assuming the following $(2\times 2)$-matrix
representation:
\begin{equation}
\ba{l}
A_1=\pmatrix{1/4 & 0\cr 0 & -1/4}, \quad
A_3=\pmatrix{0 & 1\cr 0&0},\\[5mm]
A_0=\pmatrix{\lambda/4&-2\cr 0&-\lambda/4}, \quad
A_2=\pmatrix{-\lambda^2/8 & \lambda\cr 0&\lambda^2/8}.
\ea
\ee
Thereby from (46), (49) and (55) we obtain the next reduced curvature 1-form
$\Gamma\in \Lambda^1(M)\otimes\cal G$
\begin{equation}
\Gamma=(A_0+uA_1)dx+((u_x+u^2/2)A_1-uA_3+A_2)dt,
\end{equation}
generating parallel transporting of vectors from the representation space
$Y$ of the holonomy Lie algebra ${\cal G}(h)$:
\begin{equation}
dy+\Gamma y=0
\end{equation}
upon the integral submanifold $\bar M\subset M^4$ of the ideal ${\cal I}
(\alpha)$, generated by the set of 2-forms (44). The result (57) means also
that the dynamical system (43) is endowed with the standard Lax type
representation, having the spectral parameter $\lambda\in{\bf R}$ necessary for
its integrability in quadratures.

In the case when the condition
\[
{\cal G}(h)={\cal G}(h)_1:=span_{\bf R}\{\nabla^m_x\nabla^n_t\Omega\in{\cal G}:\,
m+n=\overline{0,1}\}
\]
is assumed satisf\/ied, one can compute that
\be
\ba{l}
{\cal G}(h)_1:=span_{\bf R}\{\nabla^m_x\nabla^n_tg_j\in{\cal G}: \,
j=\overline{1,2}, \,m+n=\overline{0,1}\}\\[2mm]
\Rightarrow\qquad span_{\bf R}\{g_j\in{\cal G};\,
\partial g_j/\partial x+[g_j,A_0+A_1u^{(0)}], \,
\\[2mm]
\partial g_j/\partial t+[g_j, u^{(1)}A_1+u^{(0)}A_1/2+[A_1,A_0]u^{(0)}+A_2]
\in{\cal G}:\,j=\overline{1,2}\}\\[2mm]
\Rightarrow \qquad span_{\bf R}\{A_1,[A_1,A_0],[[A_1,A_0],A_0],[[A_1,A_0],A_1],
\\[2mm]
\qquad [A_1,A_2],[[A_1,A_0],A_2]\in{\cal G}\}\Rightarrow span_{\bf R}
\{A_{j\ne 2}\in{\cal G}: \;j=\overline{1,7}\},
\ea
\ee
where, by def\/inition,
\begin{equation}
\ba{l}
[A_1,A_0]=A_3, \quad [A_3,A_0]=A_4, \quad [A_3,A_2]=A_7,
\\[2mm]
[A_3,A_1]=A_5, \quad [A_1,A_2]=A_6.
\ea
\ee
As a result, we have the following expansions for undetermined hidden elements
$A_0, A_2\in\cal G$
\begin{equation}
A_0:=\sum^7_{j=1, j\ne 2}q_{0j}A_j, \quad
A_2:=\sum^7_{j=1, j\ne 2}q_{2j}A_j,
\end{equation}
where $q_{0j}, q_{2j}\in{\bf R}$ are some real members to be found successfully
from conditions (58) and (59) as well as from the standard Jacobi identities.
Having found some f\/inite-dimensional representation of the Lie algebra
${\cal G}(h)={\cal G}(h)_1$ (58) and substituted it into (56), we will be in a
position to write down the parallel transportation equation (57) in a new
Lax type form useful for the study of exact solutions to the Burgers dynamical
system (43). The analogous calculations could be fulf\/illed  ef\/fectively
in cases of any other nonlinear dynamical systems [7,8], integrable by Lax on
some inf\/inite-dimensional functional spaces.

 \label{black-lp}
\end{document}